\documentclass[useAMS,usenatbib]{mn2e}

\usepackage{soul}
\usepackage{tabularx}
\usepackage{xspace}
\usepackage{aas_macros}
\usepackage{array}
\usepackage{url}
\usepackage{amsmath}
\usepackage{amssymb}
\usepackage[noabbrev, capitalise]{cleveref}
\usepackage{color,graphicx}

\usepackage [english]{babel}
\usepackage [autostyle, english = american]{csquotes}
\MakeOuterQuote{"}

\def\simlt{\mathrel{\rlap{\lower 3pt\hbox{$\sim$}}\raise 2.0pt\hbox{$<$}}}
\def\simgt{\mathrel{\rlap{\lower 3pt\hbox{$\sim$}} \raise
2.0pt\hbox{$>$}}}

\title[The T-RECS simulation II]{The Tiered Radio Extragalactic Continuum (T-RECS) simulation II: HI emission and continuum-HI cross-correlation.}
\author[Anna Bonaldi, Philippa Hartley, Tommaso Ronconi, Gianfranco De Zotti, Matteo Bonato]{Anna Bonaldi$^1$\thanks{E-mail: anna.bonaldi@skao.int}, Philippa Hartley$^1$, Tommaso Ronconi$^{2, 3,4}$, Gianfranco De Zotti$^{5,6}$, 
\newauthor
Matteo Bonato$^7$\\
1 SKA Organization, Jodrell Bank, Lower Whitington, Macclesfield, SK11 9FT, UK\\
2 SISSA, Via Bonomea 265, 34136 Trieste, Italy\\
3 INAF - Osservatorio di Astrofisica e Scienza dello Spazio (OAS), Via Gobetti 93/3, I-40127
Bologna, Italy\\
4 CN-HPC: National Centre for HPC, Big Data and Quantum Computing - spoke 3, Italy\\
5 INAF - Osservatorio Astronomico di Padova, Vicolo dell'Osservatorio 5, I-45122 Padova, Italy\\
6 Nicolaus Copernicus Academy, Poland\\
7 INAF - Istituto di Radioastronomia and Italian ALMA Regional Centre, Via Gobetti 101, I-40129, Bologna, Italy}

\begin{document}

\pagerange{\pageref{firstpage}--\pageref{lastpage}} \pubyear{2017}

\maketitle

\label{firstpage}

\begin{abstract}
In this paper we extend the Tiered Radio Extragalactic Continuum Simulation (T-RECS) to include HI emission. 
The HI T-RECS model is based on the most recent HI mass function estimates, combined with prescriptions to convert HI mass to total integrated HI flux. It further models source size, morphology and kinematics, including rotational velocity and HI line width.  
The continuum T-RECS model is updated to improve the agreement with deeper number counts available at 150\,MHz. The model for star-forming galaxies (SFGs) is also modified according to the most recent indications of a star formation rate (SFR)--radio luminosity relation, which depends primarily on stellar mass rather than redshift. 
We further introduce prescriptions to associate an HI mass to the T-RECS radio continuum SFG and Active Galactic Nuclei (AGN) populations. This gives us a way to meaningfully associate counterparts between HI and continuum catalogues, thus building HI $\times$ continuum simulated observations. Clustering properties of the sources in both HI and continuum are reproduced by associating the galaxies to dark matter haloes of a cosmological simulation. 
We deliver a set of mock catalogues, as well as the code to produce them, which can be used for simulating observations and predicting results from radio surveys with existing and forthcoming radio facilities, such as the Square Kilometre Array (SKA).

\end{abstract}
\begin{keywords}
radio lines: galaxies, radio continuum: galaxies, galaxies: luminosity function, mass function, large-scale structure of Universe, software: simulations
\end{keywords}

\section{Introduction}

The neutral gas component of galaxies can be traced by the 21-cm "spin-flip" transition of neutral hydrogen (HI), which can be detected over the radio continuum emission at a rest frequency of $\nu_{21\rm{cm}}\simeq1.420$\,GHz \citep{1970ITIM...19..200H}. 

  So far, due to the faintness of the signal, most untargeted HI observations have observed the local Universe (e.g. HIPASS, \citealt{hipass1}, \citealt{hipass2}, ALFALFA \citealt{2005AJ....130.2598G,2018ApJ...861...49H}), with a few targeted fields to extend to slightly higher redshifts (AUDS \citealt{2021MNRAS.501.4550X}, BUDHIES \citealt{2023MNRAS.519.4279G}).
  A new generation of facilities is allowing HI surveys that will greatly improve over their predecessors in terms of redshift range, survey area, sensitivity and spatial resolution: CHILES on the Jansky Very Large Array \citep[JVLA,][]{chiles}, WALLABY and DINGO on the Australian SKA Pathfinder \citep[ASKAP,][]{wallaby,dingo}, LADUMA and MIGHTEE on the SKA Precursor MeerKAT \citep{2016mks..confE...4B,mightee,2023arXiv230413051P}  and the Apertif imaging surveys \citep{2022AA...667A..38A}.  These observational efforts will culminate with the Square Kilometre Array Observatory (SKAO) planned HI observations \citep{2015aska.confE.167S}. 
  
Interstellar gas and stars, and the link between them, are at the heart of galaxy formation and galaxy evolution studies. They together account for the galaxy's baryonic content, and the former fuels the latter through bursts of star-formation activity.  
 In this context, HI and continuum surveys combined are a particularly powerful tool, with the former tracing neutral gas and the latter tracing star formation \citep[e.g.][]{2016MNRAS.460.3419M}. 
In \cite{trecsI} (hereafter, T-RECS I) we delivered a code, T-RECS\footnote{https://github.com/abonaldi/TRECS.git}, to produce mock radio continuum catalogues consistent with the most recent data and models, with the aim of predicting what upcoming and future radio continuum surveys could look like, and to optimise survey specifications to science goals. This work extends its scope to HI surveys, as well as continuum--HI combined studies.

One way to produce mock HI catalogues would be to add HI properties to the radio continuum modelling presented in T-RECS I. However, this would produce catalogues selected in radio continuum rather than HI, which would not be adequate to simulate the untargeted HI surveys mentioned above. In light of this, we have implemented two independent modules: one producing catalogues selected in HI (described in Sec. \ref{sec:himodel}) and one producing catalogues selected in radio continuum (updated from T-RECS I as described in Sec. \ref{sec:contmodel}). As a further step, those two outputs can be combined in a consistent HI $\times$ continuum catalogue set as detailed in Sec. \ref{sec:cross}.  This approach explicitly allows for sources in one catalogue either to have or not have a counterpart in the other, based on expected correlations between HI and radio continuum properties and on the respective selection functions. It also gives the option to produce either of the two simulations as stand-alone. This code architecture also easily supports extending to other wavelengths in the future (e.g. optical, infra-red).
The issue of including realistic clustering into the simulation is discussed in Sec \ref{sec:clustering}. Finally we draw our conclusions in Sec \ref{sec:conclusions}.

In this work, $H_0=67$, $\Omega_{m}=0.32$, $\Omega_\Lambda=0.68$, which are the \cite{Planck2018} $\Lambda$CDM best-fit parameters rounded to two significant figures.

\section{HI model}\label{sec:himodel}

\subsection{HI flux}
Following \cite{10.1111/j.1365-2966.2012.21987.x}, the total integrated HI flux of a source, $f_{\rm HI}$, can be directly linked to the neutral hydrogen mass, $M_{\rm HI}$. By adopting an optically-thin approximation for the HI, where self-absorption is negligible, they obtain 
\begin{equation}
    f_{\rm HI}=\frac{M_{\rm HI}}{49.8} d_{\rm L}^{-2}\,{\rm Jy\,Hz} \label{hiflux}
\end{equation}
where $d_{\rm L}$ is the luminosity distance of the source in Mpc and $M_{\rm HI}$ is the HI mass in units of $M_{\odot}$. This relation essentially converts between HI luminosity function and HI mass function and it is yet another example of the intrinsic power of HI observations for probing galaxy physics. 

To generate T-RECS HI catalogues, we combined this relation with the best-fit $z=0$ mass function from  \cite{jones2018}:  

\begin{equation}
\Phi(M_{\rm HI}) = \ln(10) \phi _{\star} (M_{\rm HI}/M_{\rm HI}^{\star})^{(\alpha+1)} \exp[-M_{HI}/M_{\rm HI}^{\star}]\label{mhi},
\end{equation}
with $\log(M_{\rm HI}^\star)$=9.94 \,$M_{\odot}$, $\phi_{\star}=4.5 \times 10^{-3}$, $\alpha=-1.25$. 

The redshift evolution of the HIMF is currently poorly constrained by the data, a situation that should drastically improve once the new-generation surveys are completed. 

\cite{2022ApJ...940L..10B} and \cite{2023arXiv230111943P} investigated the HIMF at $z\sim0.3$ with GMRT and MeerKAT data, respectively. They both found a significant evolution of both $M_{\rm HI}^\star$ and $\phi_{\star}$, which results in far fewer massive HI galaxies at $z=0.3$ than at $z=0$. 

In line with those results, we parameterize the evolution as

\begin{eqnarray}
\log(M_{\rm HI}^{\star}(z)) &=& \log(M_{\rm HI}^{\star})_{z=0} + C_{\rm evol} \times z.\\
\log \phi_{\star} (z) &=& \log \phi_{\star z=0} + \phi_{\rm evol} \times z.
\end{eqnarray}

We determine the parameters $C_{\rm evol}=-1.41$, $\phi_{\rm evol}=1.55$, by imposing that, at $z=0.32$, $\log(M_{\rm HI}^\star/M_\odot)=9.49$ and $\phi_{\star}=1.4 \times 10^{-2}$\,Mpc$^{-3}$\,dex$^{-1}$, which are the best-fit values of \cite{2023arXiv230111943P}. 

\begin{figure}
\includegraphics[width=8.5cm]{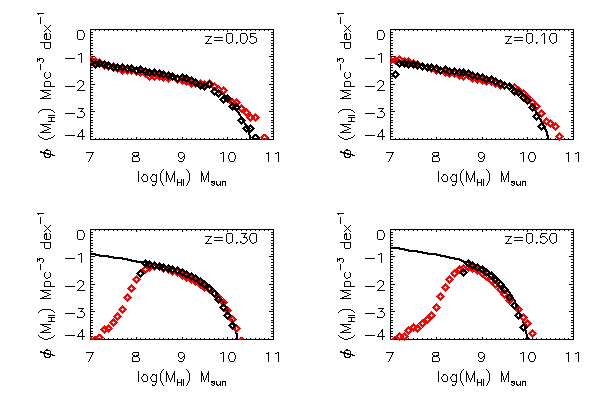}
\caption{HI mass function from Jones et al. (2018) with redshift evolution constrained by Paul et al. (2023, back lines), compared to the distribution of T-RECS HI masses for an HI catalogue of 25\,deg$^2$, and flux limit $f_{\rm HI}=1$\,Jy\,Hz (black symbols). In red we also show the distribution of the HI mass proxy obtained for a companion continuum simulation of 25\,deg$^2$, $f_{1.4 \rm GHz}\geq 100$\,nJy (see sec. \ref{sec:hiprox}).}
\label{fig:HIMF}
\end{figure}

Given the lack of data on the HIMF at higher redshifts, we restrict the redshift range of the simulation to $z=0$--0.5, which sets the frequency range for HI surveys we can simulate to 1420.40--946.9\,MHz. 
Fig.\ref{fig:HIMF} shows the adopted HI mass function (black line) compared with a T-RECS realization (black symbols) for different redshifts.

\subsection{Source size and morphology}
 
There is a tight relation between HI mass and the diameter of the HI disk $D_{\rm HI}$ in kpc, defined at an HI surface density $\Sigma_{\rm HI}$ of 1\,$M_{\rm sun}\,$pc$^{-2}$  \citep[e.g.,][]{1997A&A...324..877B, 2001A&A...370..765V,2002A&A...390..829S,2005A&A...442..137N,2014MNRAS.441.2159W,2016MNRAS.463.4052P,2021MNRAS.502.5711N},  which appears to be independent of redshift in the so far explored range \citep{2022MNRAS.512.2697R}. 
We adopt the recent relation by \cite{2021MNRAS.502.5711N}:
\begin{equation}
\log M_{\rm HI} = (1.95 \pm 0.03) \log D_{\rm HI} + (6.5 \pm 0.04),
    \label{dhi}
\end{equation}
which was derived by using a complete sample of 228 WHISP galaxies including all morphological types. This relation is in particularly good agreement with \cite{1997A&A...324..877B} and  \cite{2016MNRAS.460.2143W}, which have a similar morphological selection. It is steeper than \cite{2001A&A...370..765V,2002A&A...390..829S,2005A&A...442..137N}, which focused on specific morphological types. 

$D_{\rm HI}$, obtained by inverting eq. (\ref{dhi}), is finally converted from the physical size in kpc to an angular size in arcsec depending on the angular diameter distance of the source. For consistency with the continuum module, where the size of SFGs is in terms of a scale radius, the T-RECS catalogue contains the radius $R_{\rm HI}=D_{\rm HI}/2$.  
To further complete the morphological description of the HI sources, the catalogue contains galaxy inclination $i$ which, in the hypothesis of a random orientation, follows a $\sin(i)$ distribution.

Following e.g. \cite{1958MeLuS.136....1H}, galaxy inclination is linked to the axis ratio $b/a$
\begin{equation}
    {\rm cos}^2i = \frac{(b/a)^2-\kappa^2}{1-\kappa^2} \label{i2q}
\end{equation}
where $\kappa$ is the ratio of smallest to largest axis of an oblate spheroid which best represents the galaxy's 3-dimensional shape. Early measurements indicated values of $k=0.2$ for spirals and $k=0.5$ for ellipticals. 
\cite{2013MNRAS.434.2153R} derive updated distributions for $\kappa$ by studying the Sloan Digital Sky Survey (SDSS) Galaxy Zoo objects. A Gaussian fit to these distributions yields $\kappa=0.267 \pm 0.102$ for spirals and $\kappa=0.438\pm 0.196$ for ellipticals. We use the previous relations and the galaxy's size and inclination  to the derive major and minor axes. Since HI-rich galaxies are predominantly of spiral morphology, we adopt for the whole HI sample the spiral distribution of \cite{2013MNRAS.434.2153R}. 

\subsection{Source HI line width}\label{sec:kinetic}
The width of the 21-cm emission line of neutral hydrogen is related to the circular velocity, corrected for the galaxy's inclination, by
\begin{equation}
w_{50}=2\sin(i)v_{\rm max}
\label{eq:w50}
\end{equation}
\citep[e.g.][]{2022MNRAS.509.3268O}, where $w_{50}$ is the full with at half maximum (FWHM) of the 21 cm line and $v_{\rm max}$ is the maximum circular velocity.

\cite{2019MNRAS.483L..98K} derive empirical scaling relations of $v_{\rm max}$ with the galaxy's baryonic mass $M_{\rm b}$ (Baryonic Tully-Fisher relation, BTFR), as well as dark matter halo mass, $M_{\rm h}$,  from 120 late-type galaxies from the SPARCS database. We use their BTFR results obtained by using a "flat" prior on the mass-to-light ratio, to derive $v_{\rm max}$ from the baryonic mass.  The latter has been modelled as $M_{\rm b}=M_{\rm HI}+M_{\rm star}+M_{\rm H2}+M_{\rm HII}$, where $M_{\rm star}$ is the stellar mass, $M_{\rm H2}$ is the mass of molecular hydrogen and $M_{\rm HII}$ is the mass of ionised hydrogen. Although there are other baryonic components, the list above contains the majority of baryons in a galaxy. 

For $M_{\rm star}$ we use the maximum likelihood $M_{\rm star}$--$M_{\rm HI}$ relation by \cite{2021MNRAS.502.5711N}. \cite{2020ApJ...902..111W} investigate the density of atomic and molecular hydrogen, $\rho_{\rm HI}(z)$ and $\rho_{\rm H2}(z)$, in galaxies as a function of redshift. 
For $M_{\rm H2}$ we use $M_{\rm H2}=M_{\rm HI}\, \rho_{\rm H2}(z)/\rho_{\rm HI}(z)$ where $\rho_{\rm H2}(z)$ and $\rho_{\rm HI}(z)$ are the best-fit relations of \cite{2020ApJ...902..111W} (their table 1). For $M_{\rm HII}$ we use the $\log (M_{\rm star})$--$\log (M_{\rm HII}/M_{\rm star})$ relation by \cite{2014MNRAS.442.2398P} (linear fit to their Figure 2). This relation is provided in the $\log M_{\rm star}=6-13\,\log M_{\odot}$ and implies an increase of the ionised hydrogen with decreasing stellar mass. To avoid $M_{\rm HII}$ to become unrealistically high for low values of $M_{\rm star}$, we cap the maximum value of $M_{\rm HII}$ to $M_{\rm HI}$. 
Once $v_{\rm max}$ has been obtained, by using the $M_{\rm h}$--$v_{\rm max}$ relation by \cite{2019MNRAS.483L..98K} we add the modelling of the dark matter mass from HI properties, $M_{\rm h,HI}$, which is used to model source clustering in Sec \ref{sec:clustering}.

\begin{figure}
\includegraphics[width=8.5cm]{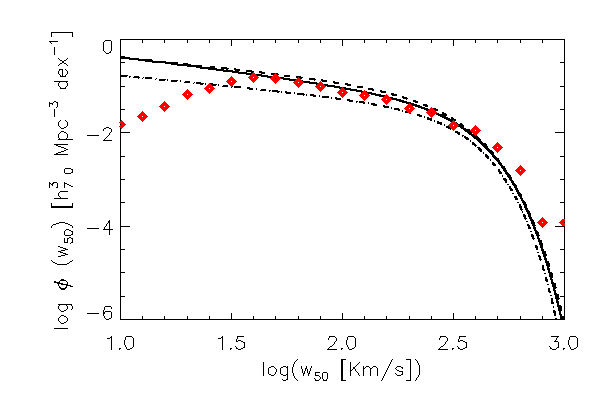}
\caption{Black lines: HIWF from \citet{2022MNRAS.509.3268O} computed from the $\alpha$.100 catalogue (solid, dashed and dot-dashed lines are for the "complete", "spring" and "fall" sky areas, respectively). Red symbols: T-RECS HIWF computed on an HI catalogue of 25\,deg$^2$, and flux limit $f_{\rm HI}=1$\,Jy\,Hz, over the 0--0.075 redshift range.}
\label{fig:HIWF}
\end{figure}

In Fig. \ref{fig:HIWF} we compare the distribution of the T-RECS $w_{50}$  at low redshifts (red symbols) with the
HI velocity width function (HIWF)  derived from the ALFALFA survey by Oman (2022; black lines).  Despite the relative simplicity of the model, there is a good agreement between the two. The decreasing number of T-RECS objects below $w_{50}\sim10^{2}$ is due to the catalogues becoming increasingly incomplete. 
When including higher redshifts, the T-RECS HIWF slowly shifts to lower $w_{50}$ values and increases in amplitude due to the evolution of the HI mass function. 
We note that our modelling of the line width relies on the presence of a plateau in the velocity curve for $v=v_{\rm flat}$. This is not representative of irregular or dwarf galaxies \citep{2009A&A...493..871S,2015AJ....149..180O} whose measured velocity width would be smaller.

\section{Radio continuum model}\label{sec:contmodel}
The radio continuum model is described in detail in T-RECS I; here we give a summary and introduce a few modifications. 
T-RECS models radio continuum sources as either star-forming galaxies (SFGs) or Radio Loud Active Galactic Nuclei (RL AGN). 
There is no explicit modelling of radio-quiet (RQ) AGN \citep[e.g.,][]{kellermann2016,Padovani2015,mancuso2017,white2017,2019MNRAS.485.3009H}. RQ AGN can be modelled as objects where star formation in the host galaxy is responsible for the radio emission. On the other hand, observations have also shown that the active nucleus can produce the radio emission. RQ AGN would contribute part of the flux of those sources that, in T-RECS, are modelled as SFGs. 

\subsection{Star-forming galaxies (SFGs)}
SFGs are modelled as in \cite{mancuso2015} as late-type, spheroidals and lensed spheroidals. The radio emission is based on redshift-dependent star-formation rate (SFR) functions and a modelling of synchrotron, free-free and thermal dust emission as a function of SFR for each of the three sub-populations. The free-free emission and the thermal dust emission are still modelled as in T-RECS I. The synchrotron emission, which is overall the dominant component, is updated in this work to  follow the most recent results. 

Across the 150\,MHz--20\,GHz frequency range, we adopt 

\begin{equation}
L_{\rm synch}(\nu)= L({\rm SFR}) \, \left(\frac{\nu}{\nu_0}\right)^{\alpha+\delta \alpha \log (\nu/\nu_0)},
\label{lsync_nu}
\end{equation}
where $L({\rm SFR})$ follows \cite{smith2021}
\begin{equation}
    \frac{L(\rm SFR)}{{\rm W/Hz}} = L_0 \left( \frac{\rm SFR}{M_{\odot}/{\rm yr}}\right)^\beta \, \left( \frac{M_{\rm star}}{10^{10}\,M_\odot} \right) ^\gamma. \label{Lsync}
\end{equation}

The frequency dependence is a power-law with spectral index that progressively steepens towards higher frequencies; the parameters are $\alpha=-0.85$, $\delta \alpha=-0.1$, $\nu_0=1.6$\,GHz. The previous relation we adopted, as in \cite{mancuso2015}, already had a steepening towards the highest frequencies. The most recent data at 150\,MHz \citep{locuss,lotss} indicate that this behaviour should be extended to the whole frequency range.  A possible mechanism for a  steepening synchrotron spectral index is synchrotron ageing. 
  
$M_{\rm star}$ in eq. (\ref{Lsync}) is the stellar mass of the galaxy. This is modelled starting from the SFR using the best-fit SFR--$M_{\rm star}$ relation of \cite{aversa2015} (their table 2, with redshift evolution). The redshift evolution of $M_{\rm star}$ induces a redshift evolution on the radio luminosity function. 
Parameters in eq. (\ref{Lsync}) are $\beta=0.850 \pm 0.005$, $\gamma=0.402 \pm 0.005$ \citep{smith2021} and $\log(L_0/[{\rm W Hz}^{-1}]) = 21.28$. The latter differs from the \cite{smith2021} value of $\log L_0/[\rm{W/Hz}]=22.10$, however our normalization is linked to that of eq. (\ref{lsync_nu}) at 150\,MHz 
$\log (L_{\rm sync}(150 {\rm MHz})/\log({\rm SFR}))=0.77$, which needs a $d\log L_0$=-0.77 correction. Once this is accounted for, there is still a difference in the overall normalization of $d\log L_0$=-0.06,W/Hz. Given that $L({\rm SFR})$ depends on both SFR and $M_{\rm star}$ as per eq. (\ref{Lsync}), differences in the distribution of either quantities between the measurements by \cite{smith2021} and the models in \cite{mancuso2015} and \cite{aversa2015} could cause some discrepancy. Furthermore, the parameters we adopt have been chosen to achieve consistency with the available data over the whole 150\,MHz--20\,GHz range rather than being optimised at 150\,MHz.

\begin{figure*}

\includegraphics[width=8.5cm]{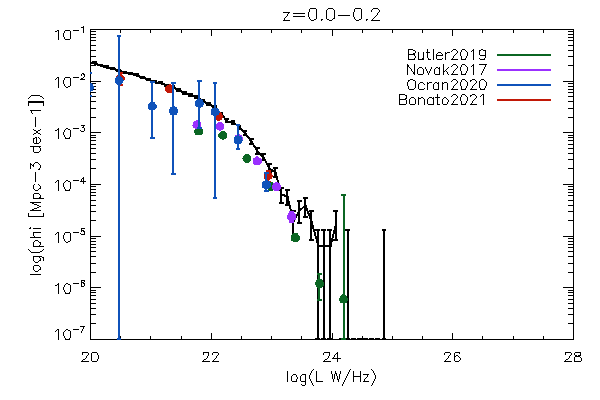}
\includegraphics[width=8.5cm]{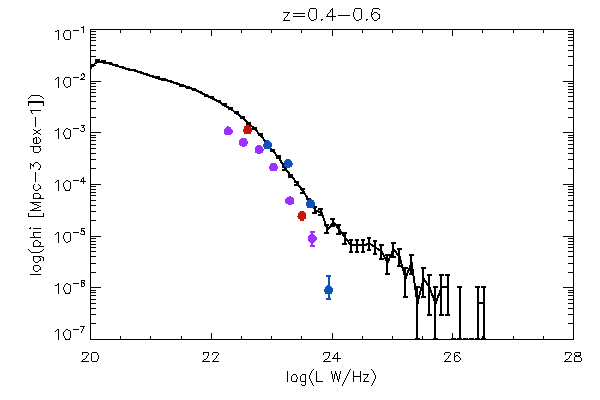}
\includegraphics[width=8.5cm]{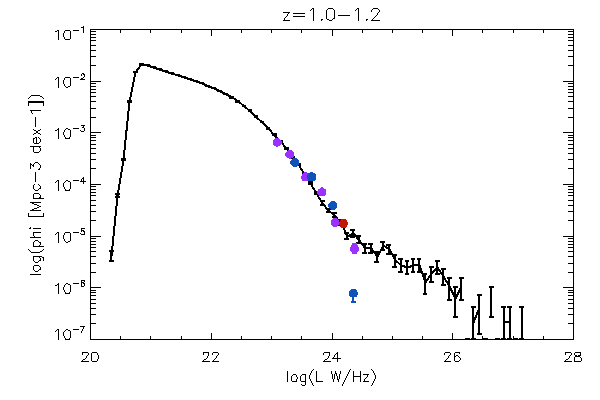}
\includegraphics[width=8.5cm]{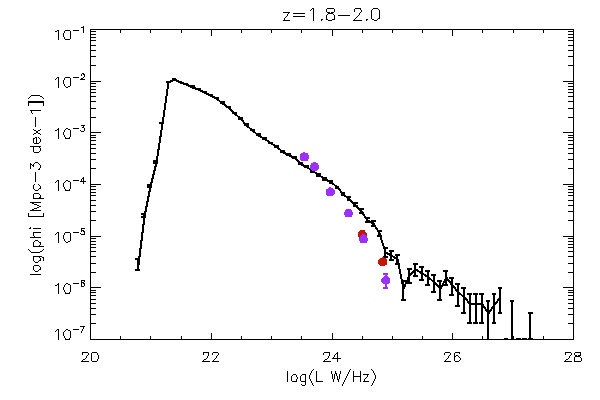}
\includegraphics[width=8.5cm]{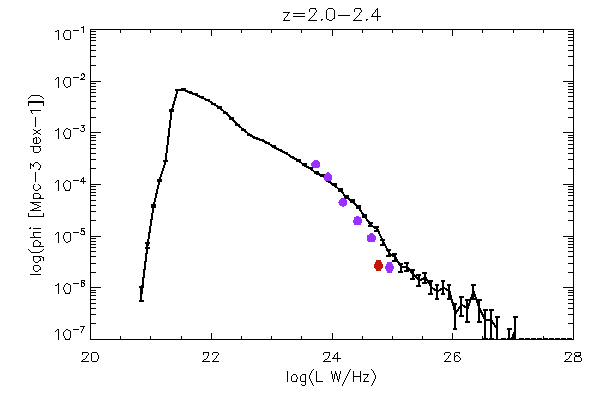}
\includegraphics[width=8.5cm]{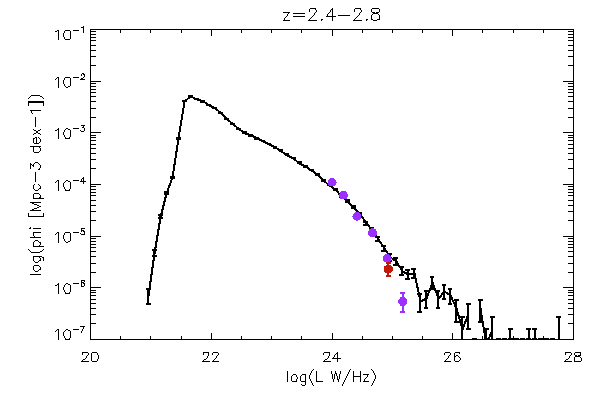}

\caption{Comparison of continuum luminosity functions between T-RECS and the available data from \citet{2017A&A...602A...5N,2019A&A...625A.111B,2020MNRAS.491.5911O,2021MNRAS.500...22B}}.
\label{fig:lums_continuum}
\end{figure*}

\begin{figure*}
\includegraphics[width=8.5cm]{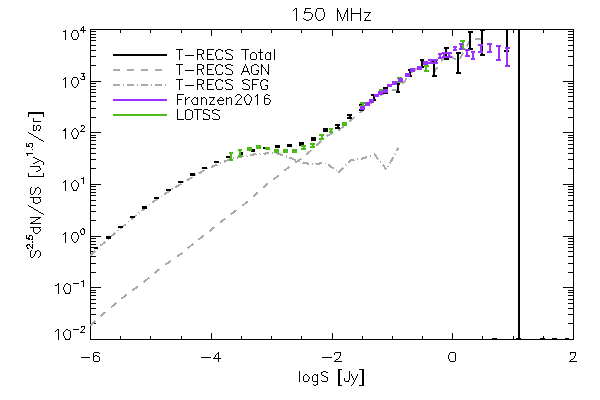}
\includegraphics[width=8.5cm]{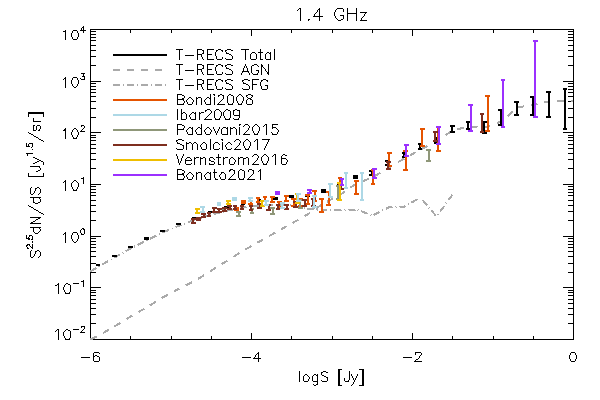}
\includegraphics[width=8.5cm]{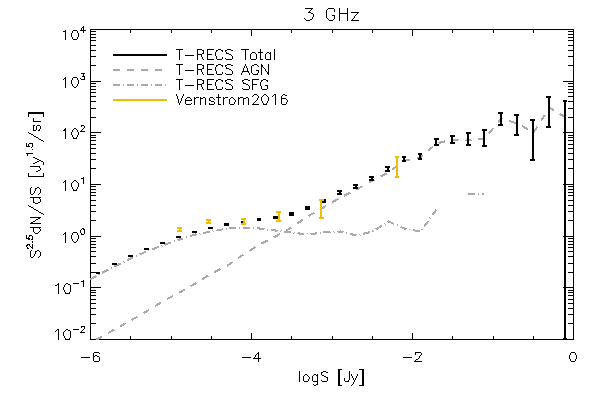}
\includegraphics[width=8.5cm]{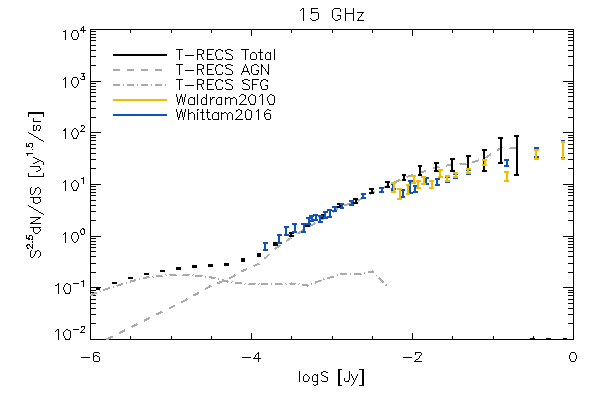}
\caption{Comparison of differential source counts in total intensity at 150\,MHz (top left) 1.4\,GHz (top right), 3\,GHz (bottom left) and 15\,GHz (bottom right) between T-RECS and the available data from \citet{franzen2016,Bondi2008,Vernstrom2016,Smolcic2017a,Padovani2015,ibar2009,whittam2016,waldram2010,davies2011,lotss,2021MNRAS.500...22B}}.
\label{fig:counts_continuum}
\end{figure*}

In Fig. \ref{fig:lums_continuum} we verify that the updated luminosity functions at 1.4\,GHz are in good agreement with the data.
The updated comparison of differential source counts in the 150\,MH--20\,GHz range is presented in Fig. \ref{fig:counts_continuum}. 

As in T-RECS I, the size and shape of SFGs is modelled in terms of elliptical Sersic profiles. As a refinement, we link galaxy inclination to galaxy ellipticity using again eq. (\ref{i2q}) with $\kappa=0.267 \pm 0.102$ for spirals and $\kappa=0.438 \pm 0.196$ for ellipticals. We make a direct association with the T-RECS radio continuum sub-populations as follows: late-type to spirals, spheroids and lensed spheroids to ellipticals \citep{2013MNRAS.434.2153R}. 

\subsection{Radio Loud (RL) AGN}
RL AGN are based on the \cite{massardi2010} evolutionary model as updated by \cite{Bonato2017}. This model classifies the AGN into steep-spectrum sources (SS-AGN), flat-spectrum radio quasars (FSRQs) and BL Lacs, with different evolutionary properties. 

The basis of our AGN model is given by luminosity functions at 1.4\,GHz for these populations and for redshift intervals ranging from $z=0$ to $z=8$. These trivially translate to flux density at 1.4\,GHz given the redshift of the source and the adopted cosmology.

To compute flux densities at other frequencies, \cite{Bonato2017} adopt power-law spectra $S \propto \nu^\alpha$ with $\alpha_{\rm FSRQ}=\alpha_{\rm BLLac}=-0.1$, and
$\alpha_{\rm steep} = -0.8$. T-RECS I modified this simple recipe by allowing for a scatter in the spectral index between different sources. Moreover, it introduced systematic variations with frequency of the spectral index distributions, in order to achieve agreement with multi-frequency observations.

An effective spectral index between
the frequencies $\nu_1$ and $\nu_2$ of sources of a given population with flux density $S_1$, within $dS_1$, at $\nu_1$,
was computed by finding the flux density $S_2$ at $\nu_2$ such that
$N_1(S_1)dS_1=N_2(S_2)dS_2$. Thus $\alpha_{\rm eff}(\nu_1,\nu_2)$ is the single spectral index relating the counts at $\nu_1$ to those at $\nu_2$. This procedure was applied to the 1.4--4.8\,GHz and 4.8--20\,GHz range, using the \citet{massardi2010} model up to 5\,GHz
and the \citet{DeZotti2005} model at higher frequencies. In this work, we extend this approach to the 150\,MHz--1.4\,GHz frequency range, where we had previously applied the \cite{Bonato2017} indices plus scatter, in order to improve the agreement with the \cite{locuss} and \cite{lotss} counts at 150\,MHz. 

The morphological model for RL AGN is the same as in T-RECS I. Following the unified model of AGN, we assume the same parent population for all our AGN classes. Intrinsic sizes are drawn using \cite{dipompeo2013}; BL-Lac and FSRQ are then assigned a small viewing angle ($\theta \leq 5$\,deg) and SS-AGN a large viweing angle ($5<\theta\leq 90$\,deg), thus creating the familiar dichotomy in observed source sizes. For SS-AGN, the description is completed by  \cite{FanaroffRiley1974}'s $R_{\rm s}$ parameter, modelled following \cite{lin2010}, to provide the FRI/FRII classification. 

For consistency with the SFG catalogue, we complement the morphological description of AGN with an elliptical/spiral classification of their host galaxy. 
The issue of radio vs optical morphology was investigated by \cite{2020ApJS..247...53K}  on a sample of 32,616 spectroscopically-selected galaxies from SDSS with radio counterpart from FIRST and VLASS. For our AGN, we adopt their result on galaxies with extended radio morphology, which have an elliptical host in 98\,\% of cases.

\subsection{Dark matter mass modelling from radio continuum properties}
As in T-RECS I, we include an estimate of the dark matter (DM) mass of galaxies based on radio continuum properties $M_{\rm h,cont}$. 
The main use of this mass is to simulate the clustering properties for our sources. Special attention has been given to ensuring that the obtained T-RECS DM mass distribution matches well the mass function from the P-millennium \citep{baugh2019} cosmological simulation, which is used to simulate clustering 
(see Sec. \ref{sec:clustering} for more details). 

The DM mass modelling is the same as T-RECS I. For SFGs, 
we used a $L_{\rm SFR}$--$M_{\rm h}$ relation of the form:
\begin{equation}
    L(M_{\rm h})= N \times \Big [\Big (\frac{M_{\rm h}}{M_b}\Big )^\alpha+\Big (\frac{M_{\rm h}}{M_b}\Big )^\omega \Big ]^{-1},\label{l(m)}
\end{equation}
where $N$, $\alpha$, $\omega$ and $M_b$ are free parameters which include
redshift evolution. We start from the \cite{baugh2019} target halo mass function as a function of redshift, and we derive the parameters so that the resulting luminosity function matches as closely as possible the radio luminosity function at 1.4\,GHz from \cite{Bonato2017}. Once the parameters are derived, eq. (\ref{l(m)})  can be inverted to obtain $M_{\rm h}$ from the 1.4\,GHz luminosity. 

For RL AGN, we start again from the halo mass function of the \cite{baugh2019} simulation to draw a sample of galaxies with plausible DM mass distribution for each T-RECS redshift interval. For all galaxies in the sample, we compute the stellar mass with the \cite{aversa2015} $M_{\rm star}=F(M_{\rm h})$ relation (their Table 2, including redshift evolution). 

\cite{janssen2012} derived the fraction of galaxies hosting an RL AGN as a function of the host galaxy stellar mass, $M_{\rm star}$, separately for Low Excitation Radio Galaxies (LERGs) and High Excitation Radio Galaxies (HERGs).  We use this result to extract subsamples of HERGs and LERGs from the initial galaxy sample. We finally compute the $M_{\rm h}$ distribution of the two sub-samples, which, once appropriately normalised, can be used to draw halo masses suitable for a HERG/LERG host. We map the HERG/LERG populations into our T-RECS observational categories as follows: FSRQs to the HERG population; BL Lacs to the LERG population; SS-AGN morphologically classified as FR\,II/FR\,I to the HERG/LERG population (see TRECS I for more details).

 \subsection{HI mass modelling from radio continuum properties}\label{sec:hiprox}
 The interplay between gas and stars in the lifecycle of a galaxy results in a correlation between HI mass, stellar mass, and SFR, which is exemplified by the Kennicutt-Schmidt Law \citep{1998ApJ...498..541K}. Scaling relations between HI mass tracers and SFR tracers have been derived \citep[e.g.][just to cite some of the most recent]{2018PASP..130i4101Z,2013MNRAS.433.1425B,2010MNRAS.403..683C,2011MNRAS.415.1797C,2015ApJ...808...66J,2012ApJ...756..113H,2022ApJ...935L..13S}; the correlation is well established, with a scatter indicative of the many factors that regulate SFR. Among those, there is feedback from AGN activity, that heats the surrounding gas therefore reducing star formation. Additionally, an important role is played by galaxy type. Late-type galaxies, associated with spiral/disk morphologies, are typically HI-rich and actively forming stars, while early-type galaxies, with spheroidal/elliptical morphology, are typically HI-poor and have low star formation.

The SFR--$M_{\rm{HI}}$ correlation mentioned above provides a relatively straightforward way to add HI mass to the modelling of T-RECS star-forming galaxies. We call this HI mass $\tilde{M_{\rm HI}}$, to distinguish it from the HI mass modelled for HI sources in Sec. \ref{sec:himodel}. It can be written as:
 \begin{equation}
    \log (\tilde{M}_{\rm HI}) = A \log ({\rm SFR}) + B \label{line}
 \end{equation}
with a scatter $\sigma$; the slope and normalization $A$ and $B$ are in general redshift-dependent. 

We derive the parameters $A$ and $B$ by imposing that the resulting $\tilde{M}_{\rm HI}$ distribution is consistent with the HI mass function adopted for the HI model in Sec. \ref{sec:himodel}. 
The redshift evolution as in \cite{2023arXiv230111943P} means that the maximum HI mass changes relatively rapidly as a function of redshift. The evolution of the SFR function in \cite{mancuso2015} is not as fast, and imparts a significant redshift dependence of the parameters $A$ and $B$. We find that 
\begin{eqnarray}
A(z)&=&1.1 -2.46\,z +2.06\,z^2 \\
B(z)&=&9.5-z
\end{eqnarray}
and a scatter $\sigma \log \tilde{M}_{\rm HI}=0.2$ give a good agreement in the considered redshift range.

\begin{figure}
     \centering
     \includegraphics[width=8.5cm]{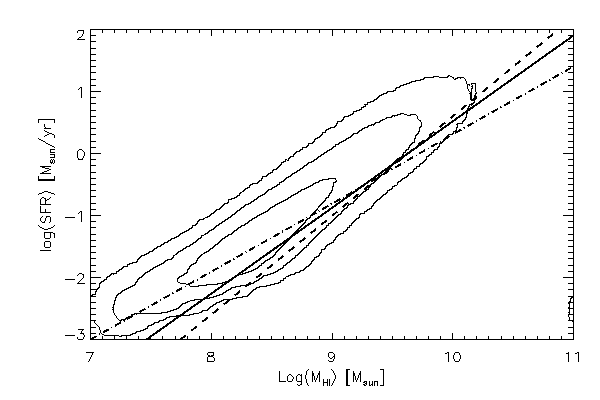}
    \caption{Relation between SFR and $\tilde{M}_{\rm HI}$ for the $z \leq 0.3$ continuum catalogue, compared to \citet{2021MNRAS.502.5711N} (dashed line), \citet{2014ApJ...782...90C} (dot-dashed line) and \citet{2015A&A...582A..78M} (solid line). Contours include are 99\%, 90\% and 50\% of all galaxies. }
         \label{fig:cross}
 \end{figure}

In Fig.\ref{fig:cross} we show the $\log (\rm SFR)$--$\log(\tilde{M}_{\rm HI})$ relation for the T-RECS catalogue up to $z=0.3$ compared to the best fits from \cite{2021MNRAS.502.5711N}, \cite{2014ApJ...782...90C} and \cite{2015A&A...582A..78M}. The slope of the correlation is consistent with the observational estimates. However, our normalization is higher, at $\Delta \log({\rm SFR})\sim 0.3$--0.4.  Using one of the observed relations instead of the abundance-matching parameters produces larger HI masses than what would be consistent with the HI mass function by \cite{jones2018} plus redshift evolution. In other words, this is essentially a tension between the \cite{mancuso2015} model of SF and the \cite{jones2018} HI mass function, which indicates that  there is still some distance to cover to get a consistent model across both observables. 

Several relations have been published between $M_{\rm HI}$ and the stellar mass, $M_{\rm star}$
\citep[e.g.][]{2010MNRAS.403..683C,2012ApJ...756..113H,2015MNRAS.447.1610M, 2021MNRAS.502.5711N}, which could be used as an HI mass proxy for the AGN population. The relation appears to be linear at the low-mass end, and to flatten at the high-mass end due to the dominance of early-type, HI-poor galaxies. We adopt the maximum likelihood relation by \cite{2021MNRAS.502.5711N} up to $M_{\rm star}=10^{10}\,M_{\odot}$ and \cite{2010MNRAS.403..683C} for higher masses. The latter was derived on an unbiased sample of galaxies selected by stellar mass and therefore should contain the gas-poor systems that are typically missed by the current untargeted HI surveys. 

In Fig. \ref{fig:HIMF} we compare the HI mass function (black line and symbols) with the HI continuum mass proxy distribution (red symbols). Overall, there is a satisfactory agreement between the two, ensured by the abundance-matching method used to derive the parameters in eq. (\ref{line}), as well as a plausible modelling of the HI mass proxy for the AGN population. 

The discrepancy at the low-mass end is due to the different selection functions. As mentioned, the selection in HI flux almost corresponds to a selection in mass, and gives high completeness of the sample down to the last mass bin. Conversely, the selection in continuum flux means that the mass proxy sample is progressively incomplete. 
Departures from the two distributions not due to completeness can be ascribed to the simplicity of eq. (\ref{line}), which cannot guarantee a perfect fit over the whole mass range.   At the high-mass end, the mass-proxy distribution tends to over-predict the number of objects. This region of the distribution contains very few objects, and therefore its weight in the abundance-matching scheme is relatively low.

 \section{Continuum $\times$ HI model} \label{sec:cross}
\begin{figure}
     \centering
     \includegraphics[width=8.5cm]{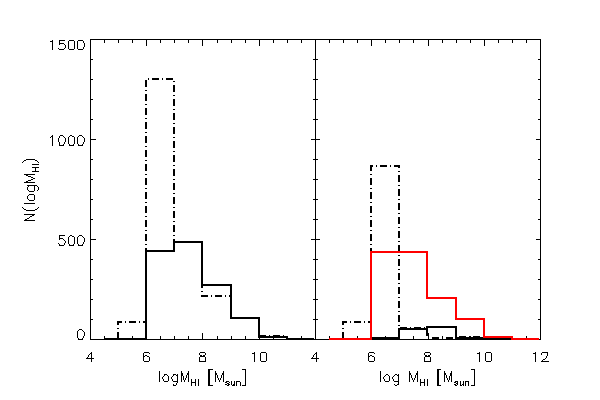}
    \caption{\emph{Left:} Distribution of $M_{\rm HI}$ for an HI catalogue of 25\,deg$^2$ $f_{\rm HI}\geq 1$\,Jy/Hz and $z=$0.475--0.525 (solid line) and of $\tilde{M}_{\rm HI}$ for a continuum catalogue with the same sky area and redshift range and $S_{\rm 1.4\,GHz}\geq 100\,$nJy (dot-dashed line). \emph{Right:} distribution of $M_{\rm HI}$ for the cross-matched catalogue (red solid line); HI unmatched sources (black solid line) and continuum un-matched sources (black dot-dashed line). }
         \label{fig:histo_mh}
 \end{figure}

\begin{figure}
     \centering
     \includegraphics[width=8.5cm]{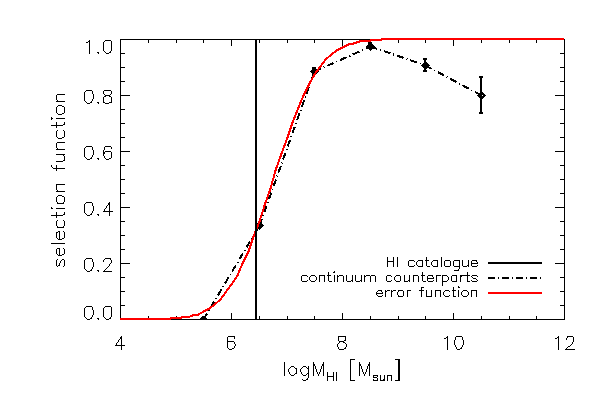}
    \caption{Selection function as a function of HI mass for the HI catalogue (black solid line) and the matched continuum counterparts (black diamonds and dot-dashed line) for the example in Fig. \ref{fig:histo_mh}. Superimposed in red is the theoretical selection function corresponding to a Gaussian error $\sigma \log M=0.65$ and an offset $\Delta \log M=0.3$, which is a good fit to the observed  selection.}
         \label{fig:mhi_select}
 \end{figure}

So far, the T-RECS HI and continuum  produced with the modules described in Secs. \ref{sec:himodel} and \ref{sec:contmodel} are independent. 
Each produces statistically plausible samples of sources, but the two samples are uncorrelated and have different sky coordinates. 

If the two catalogues were used to represent the same area of the sky, one would expect a certain number of positional cross-matches between the two, as the same galaxy is visible in both bands. In this section, we discuss how a consistent HI $\times$ continuum catalogue can be constructed by associating counterparts between the two catalogues and rewriting one of the two coordinate sets to match.

First of all, counterparts are only  explored between galaxies that belong to the same redshift slice between the two simulations. We consider a sample $\{i\}$ of continuuum sources and $\{j\}$ of HI sources, which have been selected with $S^i>S1$ and $f^j_{\rm HI}>F1$, where $S^i$ are the fluxes in continuum, $f^j_{\rm HI}$  the fluxes in HI, and $S1$, $F1$ are the respective detection limits. If we ignore redshift variation within a slice, from eq. (\ref{hiflux}) we can rewrite the HI selection in terms of HI mass, as $M^j_{\rm HI}>M1$ where $M^j_{\rm HI}$ and $M1$ are the HI masses corresponding to $f^j_{\rm HI}$ and $F1$ at the given redshift. 

The counts of both the $\{i\}$ and $\{j\}$ samples come from the appropriately normalised distributions (luminosity function, SFR function and HI mass function) and need to be conserved. Therefore, all objects are retained, and in general only some of them have a counterpart. 
Based on the HI selection function, continuum sources belonging to $\{i\}$ would be visible also in HI if their HI mass is above M1. Assessing this requires a modelling of the HI mass for the continuum sources, $\tilde{M}^i_{\rm HI}$, which we have introduced and described in Sec. \ref{sec:hiprox}.

We note that $\tilde{M}^i_{\rm HI}$ does not come from sampling an expected HI mass distribution and so selecting a subsample of potential counterparts as $\tilde{M}^i_{\rm HI}>M1$ would not automatically reproduce HI source counts and could lead to objects that are mismatched with $\{j\}$. However, if we have compatible $\tilde{M}_{\rm HI}$ and $M_{\rm HI}$ distributions, the differential counts are similar $N(M_{\rm HI}) \sim N(\tilde{M_{\rm HI}})$ and so are the number of sources above the same mass limit. In this case, assigning counterparts based on matching $M_{\rm HI}$ from the HI catalogue with $\tilde{M}_{\rm HI}$ from the continuum catalogue produces a similar selection function. 

As already discussed in Sec. \ref{sec:hiprox}, our modelling of $\tilde{M}_{\rm HI}$ for the SFGs, which are the vast majority of the continuum sources, has been explicitly designed to reproduce the HI mass function as best as possible, by using an abundance-matching technique to constrain its parameters (see also Fig 1). The mass match ensures that broadly compatible objects between the two catalogues are assigned, and it propagates, although with additional scatter, the modelled correlations between continuum and HI properties. For example, the SFR--$M_{\rm HI}$ correlation that has been modelled for the continuum catalogue shown in Fig. \ref{fig:cross} is conserved when we consider the HI $\times$ continuum catalogue. 

The HI selection $M^j_{\rm HI}>M1$ maps to a selection in $\tilde{M}^j_{\rm HI}$ which is smoother than the original step function, due to the various uncertainties introduced by the additional modelling. 

The HI mass-matching itself is performed with a nearest neighbour implementation. Initially, for each galaxy in the HI catalogue, the 20 nearest neighbours from the continuum catalogue are retained, as counterpart candidates. The HI galaxies are then reprocessed in order of decreasing HI mass, and a unique counterpart is assigned to each of them, if available,  as the best of the 20 candidates that has not been already matched. The 20 nearest neighbours have been chosen as a trade-off between computational speed and having enough samples for a good mass-matching. The reason for proceeding in inverse mass order is that high-mass objects are more rare and therefore more difficult to match accurately, but at the same time brighter and therefore more likely to have a counterpart. 
   \begin{figure}
    \centering     
     \includegraphics[width=9.cm]{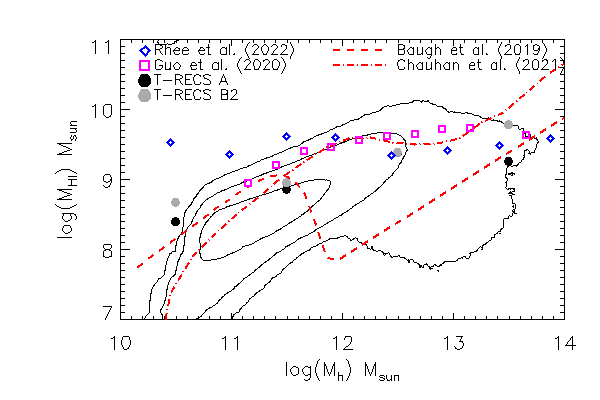}
          \caption{Median T-RECS $M_{\rm h,cont}$--$M_{\rm HI}$ (T-RECS 1) and $M_{\rm h,HI}$--$M_{\rm HI}$ (T-RECS 2) compared to HI--halo mass relation from observations \citep[][central galaxies only]{ 2020ApJ...894...92G,2023MNRAS.518.4646R} and theoretical modelling \citep{baugh2019,2021MNRAS.506.4893C}. The contours are the full distribution for the T-RECS 1 model, including 99\%, 90\% and 10\% of the galaxies.}
 \label{fig:cross_mh}
 \end{figure}

In the left panel of Fig. \ref{fig:histo_mh} we show the histograms of the HI masses $M_{\rm HI}$ for an HI catalogue with $f_{\rm HI}>1$\,Jy\,Hz and $\tilde{M}_{\rm HI}$ for a continuum catalogue with $S_{1.4\,\rm{GHz}}>$100\,nJy, over 25\,deg$^2$ and $z=0.5$.  The HI catalogue has a mass limit $\log M1=6.45$.  Similar numbers of galaxies are available in both catalogues as a function of $M_{\rm HI}$ for $\log M_{\rm HI} \geq 7$.  In the $\log M_{\rm HI}=6$ mass bin, the counts $N(M_{\rm HI})$ and $ N(\tilde{M_{\rm HI}})$ start to diverge, and 
there is an excess of sources in the continuum catalogue. 

In the right panel of  Fig. \ref{fig:histo_mh} we show the mass distribution of the cross-matched catalogue with the red line, and the un-matched sources with the black solid line and the black dash-dotted line for HI and continuum, respectively. In this case, the continuum is deeper than the HI and most HI sources have a continuum counterpart. When considering another couple of catalogues with different detection limits, the proportion of sources in the matched, HI-only and continuum-only categories would change. The $M_{\rm HI}$--$\tilde{M}_{\rm HI}$ relation for the matched objects is very tight, with $\log{M_{\rm HI}}-\log \tilde{M}_{\rm HI}=2.4 \pm 9.4 \times 10^{-3}$. 

In Fig. \ref{fig:mhi_select} we compare the selection in mass of the HI catalogue ($\log M1>6.45$) with that of the matched continuum counterparts (diamonds and dot-dashed line). The error bars are consistent with Poisson statistcs and reflect the very different number of sources that are present in the different mass bins. 

In the Gaussian approximation, for a scatter $\sigma_{M}$ on the mass $M$ and a detection limit $M1$, the selection function $F(M)$ is an error function 
\begin{equation}
F(M)=\frac{1}{2} \left[ 1-{\rm erf} \left( \frac{M_1-M}{\sigma_{M} \sqrt{2}}\right) \right].
\end{equation}

The red line in Fig. \ref{fig:mhi_select} is an error function similar to the measured selection, which corresponds to a mass limit $\log \tilde{M}_1=6.75 \pm 0.65$. The scatter is due to that of the SFR--$\tilde{M}_{\rm HI}$ relation and that of the mass matching. 
The offset $\Delta \log M=0.3$ is due to differences in the distributions of  $\tilde{M}_{\rm HI}$ and  $M_{\rm HI}$, in particular the excess of continuum sources in the $\log M_{\rm HI}=6$ mass bin. At the highest masses, the observed selection departs from the error function, with a few radio continuum galaxies not having an HI counterpart. This is due to discrepancies between the high-mass end of the HIMF and the HI mass proxy distribution (see also Fig \ref{fig:HIMF}). 

The difficulty in accurately reproducing the expected HI mass function starting from a radio continuum catalogue is one of the reasons behind our choice to model the HI independently and then cross-match with the continuum catalogue. In this way those discrepancies, rather than affecting the number of HI sources for an HI selection, only affect the number of continuum counterparts. 

We note that the method we described does not necessarily identify pairs of galaxies that are the most consistent in terms of properties other than the HI mass. To achieve this, a better approach would be to simultaneously match multiple catalogue attributes at once, for example $M_{\rm h}$, $M_{\rm star}$ and $M_{\rm HI}$. However, this approach would  increase the scatter between the matched $\tilde{M_{\rm HI}}$ and $M_{\rm HI}$ which would result in a selection of the continuum sources largely independent from the HI signal strength.

We point out that the selection functions adopted in T-RECS are an approximation of what actually achieved by real observations. In radio continuum, the ability to detect a source of a given integrated flux density depends, amongst other things, on how much of that flux is contained within the instrumental point spread function. Similarly, completeness for a given HI flux depends on the 3D size and shape of the source, particularly the line width, with sources having narrower line widths being easier to detect \citep[see, e.g.,][]{2011AJ....142..170H}. Instrumental specifications, such as spatial and spectral resolution, need to feature in an accurate description of selection functions.  Additionally,  an important role is played by the adopted source finding algorithms, which perform spatial and spectral filtering of the sources to maximise detection \citep[see, e.g.][for a comparison of different source finding methods on the same data]{2021MNRAS.500.3821B, 2023MNRAS.tmp.1348H}. If the goal is to accurately reproduce the selection functions of specific surveys, T-RECS catalogues should be generated with flux limits consistent with the deepest achievable fluxes, and further filtering of the source lists should be performed on additional source properties, as appropriate.

\section{Clustering}\label{sec:clustering}

As described in T-RECS I, the clustering properties of the sources are simulated by associating each galaxy with a dark matter sub-halo of the P-millennium \citep{baugh2019} cosmological simulation.
The cosmological parameters adopted in this simulation are:  $H_0=67.77$\,km \,s$^{-1}$\,Mpc$^{-1}$,
$\Omega_\Lambda=0.693$, $\Omega_{\rm M}$=0.307, $\sigma_8=0.8288$
\citep{planck2014}.

The reason for choosing this simulation is its very high mass resolution, which means that sub-haloes are tracked individually down to $M_{h}=1.6 \times 10^{9}\,M_{\odot}$, but also a large box size (800 comoving Mpc/h). The latter allows producing a  $5\times 5$\,deg$^2$ lightcone up to $z=8$, which  is therefore the maximum T-RECS survey size for which we can simulate clustering. This approach essentially matches galaxies to haloes by means of the dark mass $M_{\rm h}$, separately for each redshift slice, and overwrites the T-RECS latitude, longitude and redshift coordinates with those of the matched haloes. The clustering is therefore simulated in 3 dimensions. 

Two predictions for the dark matter mass of T-RECS galaxies are available: $M_{\rm h,HI}$ as modelled from the HI properties in Sec \ref{sec:himodel},  and $M_{\rm h,cont}$ as modelled from the radio continuum properties for SFGs and AGN in Sec \ref{sec:contmodel}. Given the uncertainties related with both modelling and measurement of dark matter mass, both quantities are retained as alternative models in an HI $\times$ continuum catalogue. 

Several works have explored the total HI mass in dark matter haloes (HI-halo mass relation, HIHM), which is important to understand the role of HI in galaxy formation and its connection to structure formation.  This relation has been investigated extensively in different theoretical models \citep{2014MNRAS.440.2313B,2018MNRAS.479.1627P,2019MNRAS.486.5124O,2017MNRAS.465..111K,2018ApJ...866..135V}
and observationally, via spectral stacking \citep{2020ApJ...894...92G,2023MNRAS.518.4646R}. Theoretical studies predict a positive correlation between $M_{\rm h}$ and $M_{\rm HI}$ in both the low-$M_{\rm h}$ regime, which is dominated by central galaxies, and the high-$M_{\rm h}$ regime, dominated by groups and satellite galaxies. The overall relation has a flattening at masses in between these two regimes. Observational estimates  have so far measured a much flatter relation, with $M_{\rm HI}$ only mildly depending on $M_{\rm h}$. Possible observational biases to account for these discrepancies are discussed in details in \cite{2021MNRAS.506.4893C}. 

In Fig. \ref{fig:cross_mh} we compare the $M_{\rm h,HI}$--$M_{\rm HI}$ (T-RECS B) and $M_{\rm h,cont}$--$M_{\rm HI}$ (T-RECS A) relations with some of the HIMH in the literature. 
They can only be qualitatively compared, since T-RECS's are galaxy-galaxy relations and the MIMH is the total HI content of haloes including multiple galaxies. The comparison is more meaningful at the lowest halo masses, where a single galaxy occupancy of haloes is prevalent, or when considering the central galaxies' contribution to the total HIMH. The filled circles show the median T-RECS A and T-RECS B relations;  black contours are the full T-RECS A $M_{\rm h,cont}$ model including 99\%, 90\% and 50\% of the galaxies.  Diamonds and squares are the central galaxies' observational relations by \cite{2023MNRAS.518.4646R} and \cite{2020ApJ...894...92G}. Dashed and dash-dotted lines are the theoretical relations by \cite{baugh2019} and \cite{2021MNRAS.506.4893C}. Error bars on the observational estimates have been omitted for figure clarity.

The T-RECS A model is in qualitatively good agreement with both the observational estimates and with the \cite{2021MNRAS.506.4893C} model over the mass range 
$10^{11}<M_{\rm h}<10^{13}$\,$M_{\odot}$. It also manages to reproduce the flattening of the relation towards high masses, which in T-RECS is the result of associating some HI galaxies to AGN sources, which typically have a larger halo mass. The T-RECS B model does not present any flattening and it predicts a higher HI mass for the same dark mass. This is because the relation used to model $M_{\rm h,HI}$ has been derived for late-type galaxies, which have higher HI fraction. As such, this relation fails to capture the most massive, early-type galaxies which are HI-poor. A lower DM mass for these sources would result in lower clustering. For this reason, in the clustering analysis that follows, the dark matter mass of the continuum counterpart $M_{\rm h,cont}$ has been used, whenever present. 

The implementation of the DM mass matching is based on the same nearest-neighbour approach used for the $M_{\rm HI}$ mass matching in Sec. \ref{sec:cross}. Given the very large number of DM sub-haloes contained in the lightcone, however, comparing every galaxy to every halo is a significant bottleneck in terms of execution time. At low $M_{\rm h}$ especially, there is an over-abundance of DM haloes with respect to galaxies, due to the shape of the DM mass function and selection effects for the T-RECS catalogues. On those mass ranges, and depending on the sizes of the samples to be compared, a subsample of the available sub-haloes can be randomly selected for the subsequent analysis. This has very significant advantages in terms of computational cost, at the price of a negligible increase of the scatter between the matched masses. 

\subsection{Validation for continuum sources}

Several observational estimates of clustering for AGN and SFGs are present in the literature \citep{magliocchetti2017,Hale2018,chakra2020,Bonato2017}. Details of the source selection for those studies are collected in Table \ref{tab:clustering}. 

\begin{table}
\centering
     \caption{Comparison on the selection between different radio continuum clustering estimates in the literature}
    \label{tab:clustering}

\begin{tabular}{lccl}
\hline
     Observation&$\nu$[GHz]&$S_{{\rm limit},\nu}$[mJy] &Ref. \\
    \hline
    COSMOS&3&0.013&\cite{Hale2018}\\
    COSMOS&1.4&0.15&\cite{magliocchetti2017}\\
    LH&1.4&0.15&\cite{2021MNRAS.500...22B}\\
    EN1&0.612&0.05&\cite{chakra2020}\\
    \hline
\end{tabular}
\end{table}

The observational estimates adopted the standard power-law shape for the two-point angular correlation function: $w(\theta)=A\theta^{1-\gamma}$. Since the data did not allow an accurate determination of both $A$ and $\gamma$ for each source population, they fixed $\gamma$ and fit for the normalization.

\begin{figure}
\includegraphics[width=8.5cm]{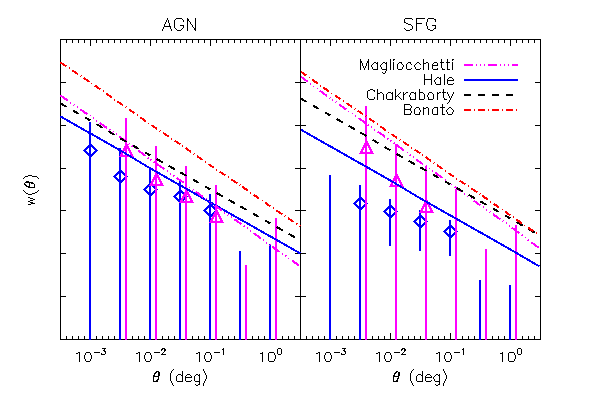}
\caption{Two-point angular correlation function $w(\theta)$ yielded by our simulation (symbols with error bars) for radio AGN (left) and SFGs (right) compared to the observational results in Table \ref{tab:clustering} (coloured lines).}
\label{fig:wtheta}
\end{figure}

To produce the T-RECS
correlation functions we use
the same flux limits of the observational estimates and a sky area of 4 $\times$4 square degrees. We use the
\cite{1993ApJ...417...19H} estimator:
\begin{equation}
w(\theta)= \frac{DD \cdot RR}{DR \cdot DR} -1, \label{hamilton}
\end{equation}
where $DD$, $RR$ and $DR$ are the number of data-data, random-random and
data-random pairs separated by $\theta$. 

The random catalogue is constructed by allocating a random position of all sources within the simulated sky area. The $w(\theta)$ is computed for 500 realizations. Each realization uses a different generation of the continuum catalogue and a different 4$\times$4\,deg$^2$ portion of the DM lightcone, which is extracted from the 5$\times$5\,deg$^2$ with a random uniform shift from the field centre in both coordinates. 

In Fig.~\ref{fig:wtheta} we show the mean observational estimates as lines of different colours. Uncertainties have not been included, in order to improve figure clarity. 
The different selection criteria are responsible for some of the differences in the measured clustering. At the same selection frequency, a deeper flux limit would correspond to less luminous sources, which are expected to be found in less massive, weakly clustered haloes. Therefore, deeper catalogues find a lower normalization of the correlation function.
A different selection frequency would favour different sub-populations, which could have different clustering properties.

T-RECS  mean values and their dispersion for catalogues with flux limits as in \cite{magliocchetti2017} and \cite{Hale2018} are shown as symbols and error bars with the same colour as the observational estimates. For the same source selection, our simulations give amplitudes of the angular correlation function consistently lower than the observational ones. 

The higher amplitudes of the observationally estimated $w(\theta)$ imply higher average bias factors, i.e. higher halo masses. As already discussed in T-RECS I, various factors can influence the comparison between the observational determinations and the T-RECS simulations. One of those is the observational classification between AGN and SFG and how this compares with the definition of the two populations in T-RECS. Another factor is the availability of high-mass haloes in the \cite{baugh2019} cosmological simulation adopted in T-RECS. The mass function of
the cosmological simulation and that inferred from the luminosity function are somewhat different, which means that there is a deficit of suitable haloes in some mass ranges and a surplus in others. Allowing for some scatter between the predicted and the associated mass alleviates the problem; however, it typically
favours association to smaller halo masses, given the
shape of the mass function.

T-RECS results are still consistent with the observations using the same flux limit, with the exception of \cite{Hale2018} SFGs. In the latter case the discrepancy is of  2.5\,$\sigma$, where $\sigma$ is the quadratic sum of errors of the observational estimates and of the simulations.

Despite the differences noted above, and taking into account the uncertainties in both determinations, the agreement between the T-RECS clustering and the empirically-determined one is reasonably good. 

\subsection{Validation for HI sources}
Observational estimates of the real-space correlation functions for galaxies selected in HI are presented by \cite{passmoor2011}, \cite{martin2012} and  \cite{papastergis2013}. Details of the survey used are reported in Table \ref{tab:clusteringHI}. 
The observational estimates adopt a power-law model for the correlation function of the form $\xi (r)=(r/r_0)^{-\gamma}$, and fit for both the $r_0$ and the $\gamma$ parameters.

We select T-RECS sources with a mass limit of $\log M_{\rm HI}=7.5\,M_\odot$ and a redshift limit of $z=0.05$, in an attempt to be comparable with the observational estimates. However, the observed samples have a complex selection function and they are neither flux- or mass-limited, which makes the comparison of the results more complicated. 

In order to get good statistics for the computation of $\xi(r)$, we use the maximum possible area for a T-RECS clustered simulation, of 25\,deg$^2$. Observational estimates were obtained within larger surveys, of $\sim$400\,deg$^2$ for ALFALFA and 29,000\,deg$^2$ for HIPASS. 

To produce T-RECS correlation functions we use again the estimator in eq. (\ref{hamilton}); we checked that using the \cite{landy} estimator as in \cite{passmoor2011}, \cite{martin2012} and \cite{papastergis2013} gives comparable results. 
For the random catalogue we use the unclustered T-RECS catalogue, which therefore has a plausible redshift distribution and random angular coordinates within the simulated sky area. The clustered catalogue contains the same sources, but the angular and redshift coordinates are replaced with those of dark haloes in the same redshift range that have similar mass. 

The $\xi(r)$ is computed for 500 realizations. Each realization uses a different generation of both the HI and continuum catalogue (which is also used in the dark mass matching process, as previously explained). 

\begin{figure}
\includegraphics[width=8.5cm]{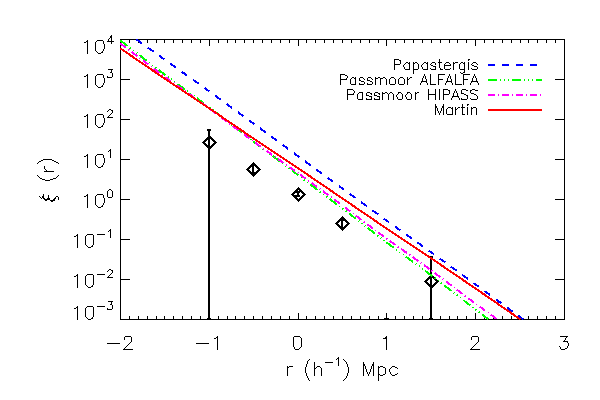}
\caption{Real-space correlation function $\xi(r)$ yielded by our simulation (symbols with error bars)  compared to the observational results in Table \ref{tab:clusteringHI} (coloured lines).}
\label{fig:clusteringHI}
\end{figure}

In Fig. \ref{fig:clusteringHI} we show the comparison between T-RECS and the observational estimates (uncertainties in the latter were not included to improve figure clarity). T-RECS results reproduce the slope of the observations and are lower in amplitude; the discrepancy is of 1.5--2.5\,$\sigma$ depending on the dataset. 

As previously mentioned, this comparison is made more difficult by the different source selection. T-RECS sources are complete down to $M_{\rm HI}=10^{7.5}$. This is the minimum mass achieved by the observational samples \citep[specifically,][]{papastergis2013}, which could in part explain the lower normalization. The HI catalogue would also most likely inherit the same issue of the continuum catalogue, that has to do with the availability of high-mass haloes in the cosmological simulation. 

\begin{table*}
\centering
     \caption{Details of the samples used for HI clustering estimates in the literature. Values of mass denoted with * are average masses rather than lowest masses.}
    \label{tab:clusteringHI}

\begin{tabular}{lcccl}
\hline
     Observation&$v_{\rm max} {\rm[km/s]}$ &$z_{\rm max}$&$M_{\rm HI,min}$ [$M_\odot$]&Ref. \\
    \hline
    ALFALFA&15,000&0.05&$10^{7.5}$&\cite{papastergis2013}\\
    ALFALFA&15,000&0.06&&\cite{martin2012}\\
    HIPASS&12,700&0.06&3.24\,$10^9$*&\cite{passmoor2011}\\
    ALFALFA&18,000&0.06&2.48\,$10^9$*&\cite{passmoor2011}\\
    \hline
\end{tabular}
\end{table*}

\section{Conclusions}\label{sec:conclusions}
This work updated the T-RECS radio continuum model presented in \cite{trecsI} to improve the agreement with the most recent observations and extended it to include HI emission from extra-galactic sources as well as the cross-correlation between the two. 

The HI model reproduces the \cite{jones2018} HI mass function and the \cite{2022MNRAS.509.3268O} HI velocity width function at $z=0$. Evolution of the HI mass function has been modelled as a linear dependence of the $\log (M^*_{\rm HI})$ and $\log \phi \star$ parameters with redshift, with amplitudes calibrated to give results consistent with \cite{2023arXiv230111943P} at $z=0.3$. Uncertainties in this modelling are due to a lack of high-redshift data and currently limit our HI simulation to a maximum redshift of 0.5. The HI model further includes source size \citep[modelled with the $\log M_{\rm HI}$--$\log D_{\rm HI}$ relation by][]{2021MNRAS.502.5711N} and morphology (galaxy type, inclination, ellipticity using \citealt{1958MeLuS.136....1H} and \citealt{2013MNRAS.434.2153R}). 

The updated radio continuum model makes use of a steepening synchrotron spectral index and the \cite{smith2021} $L({\rm SFR})$--$M_{\rm star}$ relation. The goodness of the model has been evaluated by comparing T-RECS realizations with observed luminosity functions at 1.4\,GHz as well as the most recent compilations of number counts in the 150\,MHz--15\,GHz. We also modelled an HI mass proxy for the continuum sources, $\tilde M_{\rm HI}$,  by using the SFR--$M_{\rm HI}$ correlation for star-forming galaxies and published relations between the HI mass and the stellar mass for the AGN. 

We showed that, by matching the HI mass proxy in the continuum catalogue with the HI mass in the HI catalogue, we are able to assign plausible counterparts between the otherwise independent T-RECS continuum and HI simulations. Our method ensures a correct number of counterparts by applying an HI selection function to the radio continuum sources consistent with that of the HI catalogue, within an error $\sigma_{\log M} \sim 0.65$. It also propagates modelled correlations between the radio continuum and HI properties via their link to $M_{\rm HI}$. 

The $M_{\rm h}$--$M_{\rm HI}$ relation for T-RECS galaxies is qualitatively in good agreement with the HI--halo mass relation from data \citep{2020ApJ...894...92G,2023MNRAS.518.4646R} and theoretical modelling \citep{2021MNRAS.506.4893C}.

Clustering properties were simulated by associating the galaxies with haloes of the P-millennium \citep{baugh2019} simulation of similar dark halo mass. 
We assessed the clustering properties of both the HI and the continuum catalogue and validated them against the empirical estimates of $w(\theta)$ \citep{magliocchetti2017,2021MNRAS.500...22B,chakra2020} and $\xi(r)$ \citep{passmoor2011,martin2012,papastergis2013}. The T-RECS simulation correctly reproduces the slope of both correlation functions. The normalization is in both cases lower (of 1.5--2.5\,$\sigma$). This could be explained at least in part by the difficulty in reproducing the selection functions of the observational estimates; issues with the availability of high-mass haloes in the \cite{baugh2019} dark matter simulation we derived the clustering from could also be responsible.

The code used to produce these results is available on github (https://github.com/abonaldi/TRECS.git) and can be used to produce mock observations of radio continuum and HI observations, as well as their cross-correlation. We believe these can be very useful to plan for future surveys. This code has been used as a basis of the second science data challenge \cite[SDC2][]{2023MNRAS.tmp.1348H} organised by the SKA Observatory to prepare the science community to analyse SKA HI data.  

\section*{Acknowledgements} 
We thank M. Massardi, V. Galluzzi, A. Lapi, I. Prandoni for useful suggestions on how to update the T-RECS continuum model. We thank S. Blyth for her useful comments on the paper. We also thank the anonymous referee for their review, that has led to a substantial improvement of this work. 

We acknowledge the use of computational resources at the SKA Observatory.

TR is supported by the PRIN MIUR 2017 prot. 20173ML3WW, 'Opening the ALMA window on the cosmic evolution of gas, stars and supermassive black holes', and by the Fondazione ICSC - Spoke 3 Astrophysics and Cosmos Observations - National Recovery and Resilience Plan Project ID CN-00000013 'Italian Research Center on High-Performance Computing, Big Data and Quantum Computing' - Next Generation EU.

We acknowledge usage of the Fortran \citep{backus1964fortran} and Python \citep{van2007python} programming languages, Astropy \citep{astropy2013}, NumPy \citep{harris2020numpy}, Scikit-learn \citep{pedregosa2011scikit}, LAPACK \citep{anderson1999lapack}, GSL \citep{galassi1996gsl}, HEALPix \citep{gorski2005healpix}, CFitsIO \citep{pence2010cfitsio}.

\section*{Data availability} 
The data underlying this article are available at the link https://tinyurl.com/TRECS2\footnote {The full link is https://www.dropbox.com/sh/52pbxscr8pkrdn7/AAA-RPe6QcTeQrcqEvNpSWkoa?dl=0}. The code used to generate the data is available on Gihub (https://github.com/abonaldi/TRECS.git)

\bibliography{trecs}
\bibliographystyle{mn2e_plus_arxiv}

\appendix

\section{Content of the catalogues}

The structure of each T-RECS HI and continuum catalogues is presented in Table A1 and A2, respectively. The continuum catalogues include the AGN and SFG populations. A few of the modelled quantities apply to only one of the populations and are left blank for the other. A flag value of -100 is used in this case. 

HI $\times$ continuum catalogues contain all columns in Table A1 and Table A2 with the exception of the second set of coordinates (x\_coord, y\_coord, latitude, longitude, redshift) which are rewritten after plausible counterparts between the two catalogues have been identified. Other quantities that have the same name in both catalogues are both kept, and the second set is renamed (e.g. bmaj\_1 instead of bmaj). This is because they all contain unique information, (e.g., the size as traced by the stars and by the neutral gas).

\begin{table*}
\caption{Structure of the HI catalogues released with this paper. Catalogues produced with the T-RECS code will have the same format.} 
\label{tab:HIcat}
\begin{tabular}{m{2.5cm} m{2.5cm} m{10.5cm}}
\hline
Tag Name &Units &Description\\
\hline
ID\_HI&&Numerical identifier for the source\\
MHI&log($M_{\odot}$)&HI mass\\
HI flux&mJY Hz&HI flux\\
Mh&log($M_{\odot}$)&Dark halo mass\\
Mstar&log($M_{\odot}$)&Stellar mass\\
x\_coord&degs&First angular coordinate in the flat-sky approximation  (see T-RECS I for more details)\\
y\_coord&degs&Second angular coordinate in the flat-sky approximation (see T-RECS I for more details)\\
latitude&degs&Latitude spherical coordinate for a chosen centre of the field\\
longitude&degs&Longitude spherical coordinate for a chosen centre of the field\\
redshift&&redshift\\
HI size&arcsec&Apparent size of the HI source\\
inclination&degs&galaxy inclination\\
axis ratio&&galaxy axis ratio, defined as the ratio between major and minor axis\\
bmaj&arcsec&major axis\\
bmin&arcsec&minor axis\\
PA&degs&position angle\\
OptClass&&Number identifying the optical type: 1 for elliptical, 2 for spiral\\

\hline
\end{tabular}

\end{table*}

\begin{table*}
\caption{Structure of the radio continuum catalogues released with this paper. Catalogues produced with the T-RECS code will have the same format except for the number and list of frequencies and the possibility to optionally output the luminosities for each frequency as additional columns.}
\label{tab:con_cat}
\begin{tabular}{ m{2.5cm} m{2.5cm} m{10.5cm}}
\hline
Tag Name &Units &Description\\
\hline
ID\_cont&&Numerical identifier for the source\\
Lum$_{1400}$&log(erg/s/Hz)&Luminosity at 1.4 GHz (AGN only)\\
logSFR&log($M_{\rm sun}$)/yr&SFR (SFG only)\\
I$_{\rm freq}$&mJy&Total intensity flux density of the source at frequency \emph{freq} for each of a list of $N_{\rm freqs}$ frequencies as specified by the user.\\
P$_{\rm freq}$&mJy&Polarized flux density of the source at frequency \emph{freq} for each of a list of $N_{\rm freqs}$ frequencies as specified by the user. See T-RECS I for details of the polarization model.\\
Mh&log($M_{\rm sun}$)&Dark halo mass\\
Mstar&log($M_{\rm sun}$)&Stellar mass\\
MHI\_pred&log($M_{\rm sun}$)&HI mass proxy\\
x\_coord&degs&First angular coordinate in the flat-sky approximation  (see T-RECS I for more details)\\
y\_coord&degs&Second angular coordinate in the flat-sky approximation (see T-RECS I for more details)\\
latitude&degs&Latitude spherical coordinate for a chosen centre of the field\\
longitude&degs&Longitude spherical coordinate for a chosen centre of the field\\
redshift&&redshift\\
size&arcsec&Apparent size of the source. This is the maximum size of the core+jet emission for AGN, and the scale radius of a Sersic profile for SFG.\\
inclination&degs&galaxy inclination\\
axis ratio&&galaxy axis ratio, defined as the ratio between major and minor axis\\
bmaj&arcsec&major axis\\
bmin&arcsec&minor axis\\
PA&degs&position angle\\
Rs&&Ratio between the distance between the spots and the total size of the jets, for the FR I /FR II classification (steep-spectrum AGN only)\\
RadioClass&&Number identifying the sub-population: 1 for late-type; 2 for spheroids;  3 for lensed spheroids; 4 for FSRQ, 5 for BL Lac and 6 for SS-AGN.\\
OptClass&&Number identifying the optical type: 1 for elliptical, 2 for spiral.\\
L$_{\rm freq}$&erg/s/Hz&OPTIONAL OUTPUT: Luminosity of the source at frequency \emph{freq} for each of a list of $N_{\rm freqs}$ frequencies as specified by the user.\\

\hline
\end{tabular}

\end{table*}

\section{Running the complete T-RECS pipeline}

Extending T-RECS to include HI and HI-continuum cross-correlation has meant an increase in the code complexity and an evolution in its architecture. 
Along with these modifications, we have implemented a new build system and a user interface we detail in this appendix.

When building the library, separate FORTRAN modules are compiled, each of which deals with a different modular component of the work-flow.
The work-flow is controlled by a master script, accessed as terminal command, which executes calls to the different components.
The modules called by the master script are listed below: 
\begin{itemize}
    \item continuum sampler (\texttt{trecs\_sampler\_continuum}) produces a sample of star forming galaxies and AGN to whom a dark mass and a HI mass are associated in order to allow for cross-matching with clustering and HI properties, as detailed in Sec.~\ref{sec:contmodel};
    \item HI sampler (\texttt{trecs\_sampler\_hi}) produces a sample of HI galaxies, with the total HI mass, line width, and morphological properties described in Sec.~\ref{sec:himodel}; this catalogue is still un-correlated to the one produced by the continuum module;
    \item cross-matching of HI properties with continuum properties (\texttt{trecs\_xmatch\_hi}) implements the algorithm matching continuum and HI sources, detailed in Sec.~\ref{sec:cross}.
    It is used when the outputs of the two modules have to be used as a consistent set on the same sky area.
    Since the input of this step is given by the HI and continuum catalogues, these are to be computed in advance. 
    \item cross-match of the DM-halo masses with those of the P-Millenium light-cone (Sec.~\ref{sec:clustering}) delivered with the input data-set coupled with the library (\texttt{trecs\_xmatch\_clustering}). 
    The coordinates of haloes in the P-Millenium \citep{baugh2019} light-cone embed the spatial properties of dark matter haloes and sub-haloes in a Planck cosmology.
    If for a source in the T-RECS catalogue a DM-halo counter-part is identified in the light-cone, the sky coordinates of the source are overwritten with those of the associated halo.
    This cross-match can be applied to each output catalogue of the T-RECS samplers.
    \item wrapping of the produced catalogues in just one file (\texttt{trecs\_wrapper}).
    For parallelisation and RAM-usage purposes, the T-RECS functions are launched and operate on redshift slices that are stored in separate files. The list of redshift slices is reported in table B1. This final module collates the raw outputs into a single catalogue file, rotating the field of view towards user-defined coordinates.
\end{itemize}
All of the components listed above can be run separately and are independent one from the other with some obvious exceptions (e.g. cross-matching, whether it is between continuum and HI properties or between the sampled catalogues and the DM haloes in the light-cone, requires to have first produced the relevant catalogues to match).

Running the full code requires to provide a parameter file with user-defined properties that will apply to the whole run (e.g. field of view, redshift limits, what output catalogues to cross-match).
By changing the initial random seed one can either produce different realisations or reproduce the exact results of some previous run.

As already mentioned, all of the functionalities of the library are wrapped by a command line interface that can be launched by running the 
\begin{verbatim}
$ trecs [OPTIONAL FLAGS] 
    -p|--params /path/to/parameter_file.ini 
\end{verbatim}
command.
The optional flags control the overall behaviour and which module(s) to run.
An example working call to the command would be:
\begin{verbatim}
$ trecs --continuum --HI
    --xmatch --clustering
    --wrap HI_continuum
    --params /path/to/parameter_file.ini
\end{verbatim}
the command executes continuum sampling and HI sampling then it cross-matches the two catalogues and assigns coordinates in the light-cone to as many sources as possible; finally, the raw catalogues are wrapped into one single file with sky coordinates rotated to match the user-defined field of view. 
\begin{table}
\caption{Redshift slices adopted in T-RECS}
\label{tab:zetas}
\begin{tabular}{lll}
\hline
Central $z$ & Minimum $z$ & Maximum $z$\\
\hline
0.01&0.00&0.01\\
0.02&0.01&0.03\\
0.05&0.03&0.07\\
0.10&0.07&0.12\\
0.15&0.12&0.17\\
0.20&0.17&0.22\\
0.25&0.22&0.27\\
0.30&0.27&0.32\\
0.35&0.32&0.37\\
0.40&0.37&0.42\\
0.45&0.42&0.47\\
0.50&0.47&0.52\\
0.55&0.52&0.57\\
0.60&0.57&0.62\\
0.65&0.62&0.67\\
0.70&0.67&0.72\\
0.75&0.72&0.77\\
0.80&0.77&0.82\\
0.85&0.82&0.87\\
0.90&0.87&0.92\\
0.95&0.92&0.97\\
1.00&0.97&1.10\\
1.20&1.10&1.30\\
1.40&1.30&1.50\\
1.60&1.50&1.70\\
1.80&1.70&1.90\\
2.00&1.90&2.10\\
2.20&2.10&2.30\\
2.40&2.30&2.50\\
2.60&2.50&2.70\\
2.80&2.70&2.90\\
3.00&2.90&3.10\\
3.20&3.10&3.30\\
3.40&3.30&3.50\\
3.60&3.50&3.70\\
3.80&3.70&3.90\\
4.00&3.90&4.10\\
4.20&4.10&4.30\\
4.40&4.30&4.50\\
4.60&4.50&4.70\\
4.80&4.70&4.90\\
5.00&4.90&5.10\\
5.20&5.10&5.30\\
5.40&5.30&5.50\\
5.60&5.50&5.70\\
5.80&5.70&5.90\\
6.00&5.90&6.10\\
6.20&6.10&6.30\\
6.40&6.30&6.50\\
6.60&6.50&6.70\\
6.80&6.70&6.90\\
7.00&6.90&7.10\\
7.20&7.10&7.30\\
7.40&7.30&7.50\\
7.60&7.50&7.70\\
7.80&7.70&7.90\\
8.00&7.90&8.10\\
\hline
\end{tabular}
\end{table} 

\begin{table*}
    \centering
     \caption{Optional flags of the T-RECS terminal command (\texttt{trecs}) and the different type of simulation they produce if added. The \texttt{[TAG]} argument in the last column determines which raw catalogue to wrap (e.g. \texttt{HI} for the output of the HI sampler, \texttt{HI\_continuum} for the cross-matched catalogue). The pipe symbol (\texttt{|}) in the column labels specifies that the flag can be called in two equivalent ways.}
    \label{tab:pipeline}
    \begin{tabular}{l|c|c|c|c|r}
         Simulation type & \texttt{--HI|-i} & \texttt{--continuum|-c} & \texttt{--xmatch|-x} & \texttt{--clustering|-C} & \texttt{--wrap|-w [TAG]} \\
         \hline
         HI, no clustering                      &X&&&&    \texttt{HI}\\ 
         HI, with clustering                    &X&&&X&   \texttt{HI\_clustering}\\ 
         Continuum, no clustering               &&X&&&    \texttt{continuum}\\ 
         Continuum, with clustering             &&X&&X&   \texttt{continuum\_clustering}\\
         Continuum $\times$ HI, no clustering   &X&X&X&&  \texttt{HI\_continuum}\\
         Continuum $\times$ HI, with clustering &X&X&X&X& \texttt{HI\_continuum\_clustering}\\
    \end{tabular}
\end{table*}
Tab.~\ref{tab:pipeline} summarizes the available command line options and the different types of simulations that are generated by calling the \texttt{trecs} command with different combinations of arguments.
A manual page with an explanation of all available optional flags is shown on screen by calling the \texttt{trecs} command with the \texttt{-h|--help} flag.

\bsp

\label{lastpage}

\end{document}